\documentclass[12pt]{iopart}
\usepackage[dvips]{graphicx}
\usepackage{latexsym}
\usepackage{iopams}
\usepackage{amssymb}
\usepackage{verbatim,times,bbm}

\begin{document}

\title{On the classical capacity of quantum Gaussian channels}

\author{Cosmo Lupo$^1$, Stefano Pirandola$^2$, Paolo Aniello$^{3,4}$, Stefano Mancini$^{1,5}$}

\address{$^1$ School of Science and Technology, University of Camerino, I-62032 Camerino, Italy}
\address{$^2$ Department of Computer Science, University of York, York YO10 5GH, UK}
\address{$^3$ Dipartimento di Scienze Fisiche dell'Universit\`a di Napoli
``Federico II'', Complesso Universitario di Monte Sant'Angelo, via Cintia,
I-80126 Napoli, Italy}
\address{$^4$ INFN -- Sezione di Napoli, Complesso Universitario di Monte Sant'Angelo, via Cintia,
I-80126 Napoli, Italy}
\address{$^5$ INFN -- Sezione di Perugia, I-06123 Perugia, Italy}

\eads{\mailto{cosmo.lupo@unicam.it}, \mailto{pirs@cs.york.ac.uk}, \mailto{paolo.aniello@na.infn.it}, \mailto{stefano.mancini@unicam.it}}

\begin{abstract}
The set of quantum Gaussian channels acting on one bosonic mode can be
classified according to the action of the group of Gaussian unitaries.
We look for bounds on the classical capacity for channels belonging
to such a classification.
Lower bounds can be efficiently calculated by restricting to Gaussian 
encodings, for which we provide analytical expressions.
\end{abstract}

\vspace{4mm}


\section{Introduction}

Gaussian processes are ubiquitous in physics, mathematics and information theory \cite{Ref1}.
In information theory, Gaussian channels are used to model noisy communication
lines where the noise is described as a Gaussian process.
A central problem in this field is to determine the capacity of the communication
line, that is, the maximum rate at which information can be reliably transmitted
via the noisy channel, with asymptotically vanishing probability of error \cite{InfTh}.
The seminal work of Shannon, besides founding the theory of information, has
provided the expression for the capacity of the Gaussian channel \cite{Shannon}.
On the other hand, the new research field of quantum information
science has represented a fertile extension of information theory, with a heritage
of open problems, challenges, and physical insights into the foundations and modern
applications of quantum physics \cite{QuInfTh,Capacities}.

Similarly to its classical counterpart, a quantum Gaussian channel is a mathematical
model for a communication line associated with a noise process with Gaussian characteristic \cite{QGauss}
(see also \cite{M1}, where this type of maps were considered in view of dynamical 
invariants \cite{M2} for quantum systems with quadratic Hamiltonians).
Actually, differently from the classical case, there are several nonequivalent notions
of capacities for a quantum channel, depending on whether the aim of the communication
protocol is to transfer classical or quantum information, and on whether entanglement
is used to assist the transmission \cite{Capacities}.
Physically, the model of quantum Gaussian channel can be applied in the context of
continuous variable quantum systems, that is, to the case of bosonic systems, to
model linear attenuation and amplification processes.
Typical realizations include the quantum electromagnetic field and atomic ensembles.

In the present contribution we consider the problem of evaluating the classical
capacity, that is, the maximum rate of reliable transmission of classical information,
through a quantum communication line modeled as a quantum Gaussian channel.
This problem is formulated as an optimization problem, where a suitable entropic
function --- the Holevo information --- has to be maximized over all possible protocols
to encode classical information into an ensemble of quantum states \cite{IHolevo}.
The main technical difficulty is due to the fact that the Holevo function might be
non additive \cite{superadd}, which means that the optimization has to be carried
over the whole unbounded set of encoding protocols, including those defined by ensembles
of entangled quantum states.
While the additivity of the Holevo function has been proved for a class of discrete quantum
channels \cite{Amosov}, nothing is known in general about continuous ones 
(to which the quantum Gaussian channels belong to).
Then, the problem is dramatically simplified by restricting to the one-shot capacity, i.e.,
considering only ensembles of separable states.
Actually it is further simplified if we also accept the conjecture that the optimal encoding 
ensemble for a Gaussian channel is solely made of Gaussian states \cite{Ref2}.
Here, resorting on these assumptions, we compute lower bounds on the
classical capacity of quantum Gaussian channels.


The paper proceeds as follows. In Sec.\ \ref{Gaussian} we briefly introduce the
notion and the basic properties of quantum Gaussian channels; in Sec.\ \ref{CCap}
we considered the problem of estimating the capacity of Gaussian channels and
compute lower bounds on the capacity of one-mode Gaussian channels; we end with
conclusions in Sec.\ \ref{end}.

\section{Gaussian states and channels}\label{Gaussian}

The most natural way to introduce the notion of quantum Gaussian channels is via
that of Gaussian states.
We consider the case of one-mode Gaussian channels, whose input and output
quantum systems are described by a canonical pair
$\mathbf{\hat R} = ( \hat Q , \hat P )$, obeying canonical commutation relations
$[ \hat Q, \hat P ] = i$ (here and in the following we set $\hbar=1$).
A state of the quantum system is described by a density operator $\hat\rho$,
from which one writes the characteristic function
\begin{equation}
\phi(\mathbf{Z}) = \mathrm{Tr}\left[ \hat\rho \, \hat W(\mathbf{Z}) \right] \, ,
\end{equation}
where $\mathbf{Z}=(X,Y)$ is a two-dimensional real vector, and
$\hat W(\mathbf{Z}) = e^{ i \mathbf{Z}\mathbf{\hat R}^\mathsf{T} }$ is the Weyl operator.
Gaussian states are, by definition, those with Gaussian characteristic function, i.e.,
\begin{equation}
\phi(\mathbf{Z}) = \exp{\left( i\mathbf{Z}_0\mathbf{Z}^\mathsf{T}-\frac{1}{2}\mathbf{Z}\mathbb{V}\mathbf{Z}^\mathsf{T} \right)} \, ,
\end{equation}
where
\begin{equation}
\mathbf{Z}_0 = \mathrm{Tr} \left[ \hat\rho \, \mathbf{\hat R} \right]
\end{equation}
is the vector of first-moment of the canonical variables $\hat Q$, $\hat P$, and
\begin{eqnarray}
\mathbb{V} = \mathrm{Tr} \left[ \hat\rho \left(\begin{array}{cc} \hat Q^2 & \frac{\hat Q \hat P + \hat P \hat Q}{2} \\ \frac{\hat Q \hat P + \hat P \hat Q}{2} & \hat P^2 \end{array}\right) \right] - \mathbf{Z}_0\mathbf{Z}_0^\mathsf{T}
\end{eqnarray}
is the covariance matrix (CM).
In addition to be positive definite, $\mathbb{V} > 0$, the uncertainty relation imposes the following
constraint on the CM associated with Gaussian states \cite{5p}:
\begin{equation}\label{UncRel}
\mathrm{det}( \mathbb{V} ) \geqslant \frac{1}{4} \, .
\end{equation}
The latter condition can be generalized to the case of multimode
Gaussian states, in terms of the symplectic invariants of the CM \cite{5p}.
Actually, in the case of one-mode Gaussian states, the determinant
of the CM is the only symplectic invariant \cite{symplectic}.
In the following we will evaluate the capacity of Gaussian channels
in terms of entropic functions.
We hence recall that the von Neumann entropy, $S[\hat\rho] = -\mathrm{Tr}(\hat\rho\log_2{\hat\rho})$
(measured in qubits), of a Gaussian state is a function of the
symplectic invariants of its CM.
For the one-mode Gaussian states we have
\begin{equation}\label{GEntropy}
S[\hat\rho] = h\left(\sqrt{\mathrm{det}(\mathbb{V})}\right) = \mathsf{S}\left[\mathbb{V}\right] \, ,
\end{equation}
where we have introduced the function $h(x) = (x+1/2) \log_2{(x+1/2)} - (x-1/2) \log_2{(x-1/2)}$ \cite{5pp}.

Quantum Gaussian channels are those quantum dynamical maps \cite{Capacities,dynamical}
which transform Gaussian states into Gaussian states \cite{QGauss}.
It follows that a Gaussian channel is identified by its action on
the CM and on the vector of first moments.
The preservation of the Gaussianity of the characteristic function
implies that a Gaussian channel is identified by a triplet
$(\mathbf{d}, \mathbb{T}, \mathbb{N})$, whose action on the vector
of first moment and the CM is as follows:
\begin{eqnarray}
\mathbf{Z}_0 & \mapsto & \mathbb{T} \mathbf{Z}_0 + \mathbf{d} \, , \\
\mathbb{V} & \mapsto & \mathbb{T} \mathbb{V} \mathbb{T}^\mathsf{T} + \mathbb{N} \, ,
\end{eqnarray}
where $\mathbb{N}$ is a $2 \times 2$ symmetric, positive semi-definite, matrix,
$\mathbb{T}$ is a $2 \times 2$ real matrix, and $\mathbf{d}$ is a $2$-component
real vector.
Moreover, the condition of complete positivity on the dynamical map is
characterized by the inequality --- stated in terms of the symplectic invariants
of the matrices $\mathbb{T}$, $\mathbb{N}$ ---
\begin{equation}
\mathrm{det}(\mathbb{N}) \geqslant \left[ \frac{\mathrm{det}(\mathbb{T})-1}{2} \right]^2 \, .
\end{equation}

Among the set of Gaussian channels, a remarkable role is played
by Gaussian unitary transformations.
These are characterized by triplets of the form $(\mathbf{f}, \mathbb{S}, \mathbb{O})$,
where $\mathbf{f}$ is a vector, $\mathbb{S}$ is a symplectic matrix, and $\mathbb{O}$ 
denotes the null matrix.
The composition, from the left and from the right, of the Gaussian channel
with the Gaussian unitaries $(\mathbf{f}_A, \mathbb{S}_A, \mathbb{O})$,
$(\mathbf{f}_B, \mathbb{S}_B, \mathbb{O})$, transforms the associated triplet
according to
\begin{eqnarray}
\mathbf{d} & \mapsto & \mathbb{S}_B(\mathbb{T}\mathbf{f}_A+\mathbf{d}) + \mathbf{f}_B \, , \\
\mathbb{T} & \mapsto & \mathbb{S}_B\mathbb{T}\mathbb{S}_A \, , \\
\mathbb{N} & \mapsto & \mathbb{S}_B\mathbb{N}\mathbb{S}_B^\mathsf{T} \, .
\end{eqnarray}

From the point of view of information theory, the composition from the left
and from the right of a Gaussian channel with Gaussian unitaries corresponds
to unitary pre-processing and post-processing of the channel.
The Gaussian channels can be classified according to equivalence up to
Gaussian unitary transformations: Two Gaussian channels are said to be
equivalent if there exist Gaussian unitary pre-processing and post-processing
mapping one channel into the other.
For one-mode Gaussian channels, different equivalence classes can be
identified \cite{EClasses}, which are summarized in Table \ref{tab:classes}.
For each class, one can pick up a representative Gaussian channel associated
with a triplet in the canonical form $(\mathbf{0}, \mathbb{T}_c, \mathbb{N}_c)$,
where $\mathbf{0}=\mathbb{S}_B(\mathbb{T}\mathbf{f}_A+\mathbf{d}) + \mathbf{f}_B$,
$\mathbb{T}_c=\mathbb{S}_B\mathbb{T}\mathbb{S}_A$,
$\mathbb{N}_c=\mathbb{S}_B\mathbb{N}\mathbb{S}_B^\mathsf{T}$,
and both matrices $\mathbb{T}_c$, $\mathbb{N}_c$ are diagonal
(see Table \ref{tab:classes} for details).
Besides the rank-deficient classes $\mathcal{A}_2$, $\mathcal{B}_1$, each class is identified by
the range of the parameter $\tau:=\mathrm{det}(\mathbb{T})$.
Within each class, a canonical form is characterized by the values of
$\tau$ and of the non-negative parameter $\bar n$.
The canonical form can be physically represented in terms of two-mode Gaussian
unitary transformations, where an ancillary mode is introduced in a thermal state \cite{EClasses}:
The class $\mathcal{C}$ describes linear attenuation (for $\tau < 1$) and
amplification (for $\tau > 1$) processes, which are commonly
modeled in terms of beam-splitters and linear amplifiers;
The class $\mathcal{D}$ is the conjugate channel of the linear amplifier;
Class $\mathcal{A}_1$ models the erasure process, and can be represented as the limit
of the attenuating channel for $\tau \to 0$.
Class $\mathcal{B}_2$ models the action of classical noise, its 
Stinespring dilation can be obtained with the help of two ancillary modes \cite{Eclasses}.
Finally, the classes $\mathcal{A}_2$, $\mathcal{B}_1$ present special features,
they are characterized by rank deficient canonical matrices and
can be obtained as singular limits of the previous cases.

\begin{table}[t]
\centering
\begin{tabular}{c|c|c|c}
Class & $\mathbb{T}_{c}$ & $\mathbb{N}_{c}$ & Range of $\tau =\det(\mathbb{T})$ \\
\hline
$\mathcal{A}_{1}$ & $\mathbb{O}$ & $(\bar{n}+1/2)\mathbb{I}$ & $\{0\}$ \\
$\mathcal{A}_{2}$ & $\frac{\mathbb{I}+\mathbb{Z}}{2}$ & $(\bar{n}+1/2)\mathbb{I}$ & $\{0\}$ \\
$\mathcal{B}_{1}$ & $\mathbb{I}$ & $\frac{\mathbb{I}+\mathbb{Z}}{2}$ & $\{1\}$ \\
$\mathcal{B}_{2}$ & $\mathbb{I}$ & $\bar{n}\mathbb{I}$ & $\{1\}$ \\
$\mathcal{C}$~(Att) & $~~\sqrt{\tau }\mathbb{I~~}$ & $~~(1-\tau )(\bar{n}+1/2)\mathbb{I~~}$ & $(0,1)$ \\
$\mathcal{C}$~(Amp) & $~~\sqrt{\tau }\mathbb{I~~}$ & $~~(\tau -1)(\bar{n}+1/2)\mathbb{I~~}$ & $(1,\infty )$ \\
$\mathcal{D}$ & $~~\sqrt{-\tau }\mathbb{Z~~}$ & $~~(1-\tau )(\bar{n}+1/2)\mathbb{I~~}$ & $(-\infty ,0)$%
\end{tabular}
\caption{Equivalence classes of one-mode Gaussian channels.
Each class is identified by the matrices $\mathbb{T}_c$, $\mathbb{N}_c$.
$\mathbb{I}$ and $\mathbb{O}$ respectively denote the identity and null matrices,
$\mathbb{Z}=\mathrm{diag}(1,-1)$.} \label{tab:classes}
\end{table}

\section{Classical capacity of one-mode Gaussian channels}\label{CCap}

An encoding procedure is identified by a map $\mathbf{x} \to \hat\rho_\mathbf{x}$,
which assigns a quantum state to each value of a stochastic variable, distributed
according to a probability density $p_\mathbf{x}$.

If the quantum states are subjected to the action of a quantum channel
$\hat\mathcal{E}$, the receiver has to extract information from an ensemble
composed of the states $\hat\mathcal{E}(\hat\rho_\mathbf{x})$.
According to \cite{IHolevo}, the maximum information, measured in bits,
that can be extracted from the enseble is given by the Holevo information:
\begin{equation}\label{chi}
\chi = S\left[ \int d\mathbf{x} p_\mathbf{x} \hat\mathcal{E}(\hat\rho_\mathbf{x}) \right] - \int d\mathbf{x} p_\mathbf{x} S\left[ \hat\mathcal{E}(\hat\rho_\mathbf{x}) \right] \, ,
\end{equation}
where $S$ is the von Neumann entropy.

The capacity of a quantum Gaussian channel $\hat\mathcal{E}$, associated with the
triplet $(\mathbf{d},\mathbb{T},\mathbb{N})$, is the supremum of the quantity (\ref{chi})
over all possible encoding procedures.
Among them, we consider Gaussian encodings, in which the value $\mathbf{x}$
of the stochastic variable is encoded in Gaussian states with CM $\mathbb{V}$
and first-moment $\mathbf{d}=\mathbf{x}$.
Furthermore, we assume a Gaussian form for the distribution $p_\mathbf{x}$,
with zero mean and CM $\mathbb{M}$, which is symmetric and positive semi-definite.
In these settings, we have that
\begin{equation}
S\left[ \hat\mathcal{E}(\hat\rho_\mathbf{x}) \right] =  \mathsf{S}\left[ \mathbb{T}\mathbb{V}\mathbb{T}^\mathsf{T}+\mathbb{N} \right] \, ,
\end{equation}
\begin{equation}
S\left[ \int d\mathbf{x} p_\mathbf{x} \hat\mathcal{E}(\hat\rho_\mathbf{x}) \right]  =  \mathsf{S}\left[ \mathbb{T}(\mathbb{V}+\mathbb{M})\mathbb{T}^\mathsf{T}+\mathbb{N} \right] \, .
\end{equation}
However, the maximization of the Holevo information, even when
restricted to Gaussian encodings, can lead to an infinite value of the
channel capacity.
This is due to the fact that the relevant Hilbert space is infinite-dimensional
(and, hence, it can in principle store an infinite amount of classical information),
and that the manifold of Gaussian states is not compact.
An effective cutoff in the Hilbert space can be introduced by imposing a constraint
on the average mean energy involved in the encoding process \footnote{Clearly, other
choices are possible, e.g., to constraint the output energy. However, since the Holevo
information (\ref{chi}) is obtained by optimizing over all the possible measurement
on the channel output, we do not assume any constraint on the output of the channel.}:
\begin{equation}\label{Econstraint}
\frac{1}{2} \mathrm{Tr}(\mathbb{V}+\mathbb{M}) \leqslant E \, .
\end{equation}

The optimization over Gaussian encodings furnishes a lower bound on the
capacity of the Gaussian channel:
\begin{equation}\label{lower}
\underline{C} = \max_{\mathbb{V},\mathbb{M}} \left\{ \mathsf{S}\left[ \mathbb{T}(\mathbb{V}+\mathbb{M})\mathbb{T}^\mathsf{T}+\mathbb{N} \right] - \mathsf{S}\left[ \mathbb{T}\mathbb{V}\mathbb{T}^\mathsf{T}+\mathbb{N} \right] \right\} \, ,
\end{equation}
where the maximum is over the CMs $\mathbb{V}$, $\mathbb{M}$ subjected to the
uncertainty relation (\ref{UncRel}) and to the energy constrain (\ref{Econstraint}).
In the following the maximization problem will be solved by using the Karush-Kuhn-Tucker
(KKT) method \cite{InfTh}, which generalizes the Lagrange method to the case of constraints
expressed by inequalities.

Before proceeding with the evaluation of the lower bounds on the
capacity for the Gaussian channels belonging to different
equivalence classes, we remark that the von Neumann entropy
is invariant under symplectic transformations, that is,
\begin{equation}
\mathsf{S}\left[ \mathbb{T}\mathbb{V}\mathbb{T}^\mathsf{T}+\mathbb{N} \right] = \mathsf{S}\left[ \mathbb{S}_B\mathbb{T}\mathbb{V}\mathbb{T}^\mathsf{T}\mathbb{S}_B^\mathsf{T}+\mathbb{S}_B\mathbb{N}\mathbb{S}_B^\mathsf{T} \right] \, ,
\end{equation}
for any symplectic matrix $\mathbb{S}_B$, which is interpreted as
a post-processing Gaussian unitary transformation.
It follows that the function (\ref{lower})
can be rewritten as follows
\begin{equation}
\underline{C} = \max_{\mathbb{V},\mathbb{M}} \left\{ \mathsf{S}\left[ \mathbb{T}_c\mathbb{S}_A(\mathbb{V}+\mathbb{M})\mathbb{S}_A^\mathsf{T}\mathbb{T}_c^\mathsf{T}+\mathbb{N}_c \right] - \mathsf{S}\left[ \mathbb{T}_c\mathbb{S}_A\mathbb{V}\mathbb{S}_A^\mathsf{T}\mathbb{T}_c^\mathsf{T}+\mathbb{N}_c \right] \right\} \, . \label{lower1}
\end{equation}
This expression is written in terms of the canonical matrices and the pre-processing
symplectic matrix $\mathbb{S}_A$. The latter can be further decomposed according to
Euler decomposition:
\begin{equation}
\mathbb{S}_A = \mathbb{R}\mathbb{D}\mathbb{R}^\prime \, ,
\end{equation}
where $\mathbb{R}$, $\mathbb{R}^\prime$ are orthogonal matrices
and $\mathbb{D}=\mathrm{diag}(r^{1/2},r^{-1/2})$.
Since the orthogonal matrix $\mathbb{R}^\prime$ preserves the constrain (\ref{Econstraint}),
it can be eliminated by redefining the CMs $\mathbb{V}$, $\mathbb{M}$, yielding
\begin{equation}
\underline{C} =  \max_{\mathbb{V},\mathbb{M}} \left\{ \mathsf{S}\left[ \mathbb{T}_c\mathbb{R}\mathbb{D}(\mathbb{V}+\mathbb{M})(\mathbb{T}_c\mathbb{R}\mathbb{D})^\mathsf{T}+\mathbb{N}_c \right]
-  \mathsf{S}\left[ \mathbb{T}_c\mathbb{R}\mathbb{D}\mathbb{V}(\mathbb{T}_c\mathbb{R}\mathbb{D})^\mathsf{T}+\mathbb{N}_c \right] \right\} \, . \label{lower2}
\end{equation}

\subsection{Class $\mathcal{A}_1$: erasure channel}

Since $\mathbb{T}_{c}=\mathbb{O}$, for this class of channels we trivially have
\begin{equation}
\underline{C} = \max_{\mathbb{V},\mathbb{M}} \left\{ \mathsf{S}\left[ \mathbb{N}_c \right] - \mathsf{S}\left[ \mathbb{N}_c \right] \right\} = 0 \, .
\end{equation}
Actually, by allowing non-Gaussian encoding we can similarly prove that the
channel capacity itself vanishes.

\subsection{Class $\mathcal{A}_2$}

For the channels belonging to this class, $\mathbb{T}_c=(\mathbb{I}+\mathbb{Z})/2=\mathrm{diag}(1,0)$
and $\mathbb{N}_c=(\bar n + 1/2)\mathbb{I}$.
Since the matrix $\mathbb{T}_c\mathbb{R}\mathbb{D}$ appearing in Eq.\ (\ref{lower2})
has rank one, its singular value decomposition reads
\begin{equation}
\mathbb{T}_c\mathbb{R}\mathbb{D} = \sqrt{t} \mathbb{R}'\mathbb{T}_c\mathbb{R}'' \, ,
\end{equation}
where $\mathbb{R}'$, $\mathbb{R}''$ are orthogonal matrices, and $\sqrt{t}$ is the
(non-vanishing) singular value of $\mathbb{T}_c\mathbb{R}\mathbb{D}$.
Absorbing the orthogonal matrix $\mathbb{R}''$ into the definition of the CMs $\mathbb{V}$
and $\mathbb{M}$, we obtain the following expression for the lower bound:
\begin{equation}\label{A2}
\underline{C} = \max_{\mathbb{V},\mathbb{M}} \left\{ \mathsf{S}\left[ t \mathbb{T}_c(\mathbb{V}+\mathbb{M})\mathbb{T}_c+\mathbb{N}_c \right] - \mathsf{S}\left[ t \mathbb{T}_c\mathbb{V}\mathbb{T}_c+\mathbb{N}_c \right] \right\} \, .
\end{equation}

By introducing the parameterization $\mathbb{V}=\mathrm{diag}(s^{-1}/2,s/2)$, Eq.\ (\ref{A2})
can be rewritten as the maximum of a function of $s$:
\begin{equation}
\underline{C} = \max_{s\in [s_-,s_+]} \left\{ h(\bar\nu_{\mathcal{A}_2})-h(\nu_{\mathcal{A}_2}) \right\} \, ,
\end{equation}
where
\begin{eqnarray}
\bar\nu_{\mathcal{A}_2} & = & \sqrt{ \left[ t \left(2E+1-\frac{s}{2} \right) + \bar n + \frac{1}{2} \right] \left( \bar n + \frac{1}{2} \right)} \, , \\
\nu_{\mathcal{A}_2} & = & \sqrt{ \left( t \frac{s^{-1}}{2} + \bar n + \frac{1}{2} \right) \left( \bar n + \frac{1}{2} \right)} \, ,
\end{eqnarray}
and $s_\pm = 2E+1 \pm \sqrt{(2E+1)^2-1}$.

\subsection{Class $\mathcal{B}_1$}

For the elements belonging to this class, $\mathbb{T}_c=\mathbb{I}$ and
$\mathbb{N}_c=(\mathbb{I}+\mathbb{Z})/2=\mathrm{diag}(1,0)$, Eq.\ (\ref{lower2}) reads
\begin{eqnarray}
\underline{C} & = & \max_{\mathbb{V},\mathbb{M}} \left\{ \mathsf{S}\left[ \mathbb{R}\mathbb{D}(\mathbb{V}+\mathbb{M})(\mathbb{R}\mathbb{D})^\mathsf{T}+\mathbb{N}_c \right] - \mathsf{S}\left[ \mathbb{T}_c\mathbb{R}\mathbb{D}\mathbb{V}(\mathbb{T}_c\mathbb{R}\mathbb{D})^\mathsf{T}+\mathbb{N}_c \right] \right\} \\
& = & \max_{\mathbb{V},\mathbb{M}} \left\{ \mathsf{S}\left[ \mathbb{V}+\mathbb{M}+\mathbb{D}^{-1}\mathbb{R}^\mathsf{T}\mathbb{N}_c\mathbb{R}\mathbb{D}^{-1} \right] - \mathsf{S}\left[ \mathbb{V}+\mathbb{D}^{-1}\mathbb{R}^\mathsf{T}\mathbb{N}_c\mathbb{R}^\mathsf{T}\mathbb{D}^{-1} \right] \right\} \, .
\end{eqnarray}
The matrix $\mathbb{D}^{-1}\mathbb{R}^\mathsf{T}\mathbb{N}_c\mathbb{R}\mathbb{D}^{-1}$
is symmetric with rank one, hence we can write it as follows:
\begin{equation}
\mathbb{D}^{-1}\mathbb{R}^\mathsf{T}\mathbb{N}_c\mathbb{R}\mathbb{D}^{-1} = n \mathbb{R}' \mathbb{N}_c {\mathbb{R}'}^\mathsf{T} \, ,
\end{equation}
where $n$ is its (non-vanishing) eigenvalue, and $\mathbb{R}'$ is orthogonal.
Absorbing $\mathbb{R}'$ into the definition of the CMs $\mathbb{V}$, $\mathbb{M}$,
we obtain
\begin{equation}\label{B1}
\underline{C} = \max_{\mathbb{V},\mathbb{M}} \left\{ \mathsf{S}\left[ \mathbb{V}+\mathbb{M}+n\mathbb{N}_c \right] - \mathsf{S}\left[ \mathbb{V}+n\mathbb{N}_c \right] \right\} \, .
\end{equation}

Using the parameterization $\mathbb{V}=\mathrm{diag}(s^{-1}/2,s/2)$, Eq.\ (\ref{B1}) yields
\begin{equation}
\underline{C} = \max_{s\in [s_-,s_+]} \left\{ h(\bar\nu_{\mathcal{B}_1})-h(\nu_{\mathcal{B}_1}) \right\} \, ,
\end{equation}
where
\begin{eqnarray}
\bar\nu_{\mathcal{B}_1} & = & \sqrt{ \left(\frac{s^{-1}}{2}+n\right) \left(2E+1-\frac{s^{-1}}{2}\right) } \, , \\
\nu_{\mathcal{B}_1} & = & \sqrt{\frac{1}{4}+n\frac{s}{2}} \, .
\end{eqnarray}


\subsection{Class $\mathcal{C}$: attenuating and amplifying channels}

For this class we have $\mathbb{T}_c=\sqrt{\tau} \mathbb{I}$, $\mathbb{N}_c=|1-\tau|(\bar{n}+1/2)\mathbb{I}$,
where $\tau \in (0,1)$ for the attenuating channels, and $\tau \in (1,\infty)$ for the amplifying ones.
Equation (\ref{lower2}) simplifies as follows
\begin{equation}\label{C}
\underline{C} = \max_{\mathbb{V},\mathbb{M}} \left\{ \mathsf{S}\left[ \tau\mathbb{D}(\mathbb{V}+\mathbb{M})\mathbb{D}+\mathbb{N}_c \right] - \mathsf{S}\left[ \tau\mathbb{D}\mathbb{V}\mathbb{D}+\mathbb{N}_c \right] \right\} \, .
\end{equation}

First of all, we notice that the matrices $\mathbb{D}$, $\mathbb{N}_c$
in Eq.\ (\ref{C}) are all diagonal, hence the maximum under the constraint
(\ref{Econstraint}) is reached in correspondence to diagonal CMs $\mathbb{V}$, $\mathbb{M}$.
The constrained optimization yields the analytic solution
\begin{eqnarray}
\underline{C} & = & h\left[ \tau\left(E+\frac{1}{2}\right) + |1-\tau|\left(\bar n + \frac{1}{2}\right)\frac{r+r^{-1}}{2}\right] \nonumber\\
& - & h\left[ \frac{\tau}{2} + |1-\tau|\left(\bar n + \frac{1}{2}\right) \right] \, .
\end{eqnarray}
The latter is obtained in the region of parameters for which
\begin{eqnarray}
\left( E + \frac{1}{2} \right) \pm \frac{|1-\tau|}{\tau} \frac{r-r^{-1}}{2}\left( \bar n + \frac{1}{2}\right) - \frac{r^{\mp 1}}{2} \geqslant 0 \, .
\end{eqnarray}

Outside this region the KKT method does not lead to an
analytical expression for the capacity lower bound, which can
however still be evaluated numerically by solving a transcendental
equation.
The existence of qualitative differences in the solution of the
optimization problems according to the range of the parameters has
been put forward and analyzed in details in \cite{Oleg} (see also
\cite{YS} and \cite{NJP}).

\subsection{Class $\mathcal{B}_2$: additive noise channels}

For the channels belonging to the equivalence class of the additive noise channel,
$\mathbb{T}_c=\mathbb{I}$, $\mathbb{N}_c=\bar n \mathbb{I}$, and Eq.\ (\ref{lower2})
reads
\begin{equation}\label{B2}
\underline{C} = \max_{\mathbb{V},\mathbb{M}} \left\{ \mathsf{S}\left[ \mathbb{D}(\mathbb{V}+\mathbb{M})\mathbb{D}^\mathsf{T}+\mathbb{N}_c \right] - \mathsf{S}\left[ \mathbb{D}\mathbb{V}\mathbb{D}^\mathsf{T}+\mathbb{N}_c \right] \right\} \, ,
\end{equation}
This expression can be obtained from (\ref{C}) by replacing $\tau$ with $1$, and
$|1-\tau|(\bar n + 1/2)$ with $\bar n$.
Under the conditions
\begin{eqnarray}\label{condB2}
E + \frac{1}{2} \pm \bar n \frac{r-r^{-1}}{2} - \frac{r^{\mp 1}}{2} \geqslant 0 \, ,
\end{eqnarray}
the optimization yields the following analytical expression
\begin{equation}
\underline{C} = h\left( E + \frac{1}{2} + \bar n \frac{r+r^{-1}}{2} \right) - h\left( \bar n + \frac{1}{2} \right) \, .
\end{equation}

Outside the region of parameters defined by Eq.\ (\ref{condB2})
the KKT method does not lead to a closed form for $\underline{C}$,
see also \cite{Brussels}.

\subsection{Class $\mathcal{D}$: Conjugate of the amplifying channel}

For the channels belonging to this class, $\mathbb{T}_c=\sqrt{-\tau}\mathbb{Z}$,
and $\mathbb{N}_c=(1-\tau)(\bar n + 1/2)\mathbb{I}$, where $\tau \in (-\infty,0)$
and $\mathbb{Z}=\mathrm{diag}(1,-1)$, yielding the following form for Eq.\ (\ref{lower2}):
\begin{equation}
\underline{C} = \max_{\mathbb{V},\mathbb{M}} \left\{ \mathsf{S}\left[ |\tau|\mathbb{Z}\mathbb{R}\mathbb{D}(\mathbb{V}+\mathbb{M})(\mathbb{R}\mathbb{D})^\mathsf{T}\mathbb{Z}+\mathbb{N}_c \right] - \mathsf{S}\left[ |\tau|\mathbb{Z}\mathbb{R}\mathbb{D}\mathbb{V}(\mathbb{R}\mathbb{D})^\mathsf{T}\mathbb{Z}+\mathbb{N}_c \right] \right\} \, .
\end{equation}
Since the von Neumann entropy $\mathsf{S}$ is a function of the determinant of its argument,
we can write
\begin{eqnarray}
\underline{C} & = & \max_{\mathbb{V},\mathbb{M}} \left\{ \mathsf{S}\left[ |\tau|\mathbb{R}\mathbb{D}(\mathbb{V}+\mathbb{M})(\mathbb{R}\mathbb{D})^\mathsf{T}+\mathbb{Z}\mathbb{N}_c\mathbb{Z} \right] - \mathsf{S}\left[ |\tau|\mathbb{R}\mathbb{D}\mathbb{V}(\mathbb{R}\mathbb{D})^\mathsf{T}+\mathbb{Z}\mathbb{N}_c\mathbb{Z} \right] \right\} \nonumber \\
& = & \max_{\mathbb{V},\mathbb{M}} \left\{ \mathsf{S}\left[ |\tau|\mathbb{D}(\mathbb{V}+\mathbb{M})\mathbb{D}^\mathsf{T}+\mathbb{N}_c \right] - \mathsf{S}\left[ |\tau|\mathbb{D}\mathbb{V}\mathbb{D}^\mathsf{T}+\mathbb{N}_c \right] \right\} \, . \label{D}
\end{eqnarray}
The latter expression is identical to Eq.\ (\ref{C}), upon replacing
$\tau$ with $|\tau|$, and $|1-\tau|$ with $1+|\tau|$.
Hence the solution of the maximization problem is obtained as for the class $\mathcal{C}$,
yielding
\begin{eqnarray}
\underline{C} & = & h\left[ |\tau|\left(E+\frac{1}{2}\right) + (1+|\tau|)\left(\bar n + \frac{1}{2}\right)\frac{r+r^{-1}}{2}\right] \nonumber\\
& - &  h\left[ \frac{|\tau|}{2} + (1+|\tau|)\left(\bar n + \frac{1}{2}\right) \right],
\end{eqnarray}
under the conditions
\begin{eqnarray}
\left( E + \frac{1}{2} \right) \pm \frac{(1+|\tau|)}{|\tau|} \frac{r-r^{-1}}{2}\left( \bar n + \frac{1}{2}\right) - \frac{r^{\mp 1}}{2} \geqslant 0 \, .
\end{eqnarray}

\section{Conclusions}\label{end}

We have considered the whole family of one-mode quantum Gaussian channels,
which can be classified into equivalence 
classes, and, by using Gaussian encoding, we have evaluated
lower bounds on the classical capacity under
input energy constraint.

For small enough values of the energy the optimal encodings can be evaluated numerically,
while a closed form for the lower bound has been obtained when the allowed energy
is above a certain threshold.
The presence of two different regimes for the optimization problem
has been studied in details in \cite{Oleg} (see also \cite{Brussels,YS,NJP}).
The region of parameters allowing an analytical expression for the lower bound on
the capacity includes the limit $E \to \infty$.
Using the fact that $h(x) \simeq \log_2{x}+\log_2{e}$ for $x \to \infty$,
we have the following high-energy limits of the capacity lower bounds:
\begin{eqnarray}\nonumber
\underline{C} \simeq \left\{
\begin{array}{ll}
0                                                           & \, \mbox{class} \, \mathcal{A}_{1} \\
\log_2{e}+\log_2(E)                                         & \, \mbox{class} \, \mathcal{B}_{1} \\
\log_2{e}+\log_2(E)-h(\bar{n}+1/2)                          & \, \mbox{class} \, \mathcal{B}_{2} \\
\log_2{e}+\log_2(|\tau|E)-h[|\tau|/2+|1-\tau|(\bar{n}+1/2)] & \, \mbox{classes} \, \mathcal{C}, \mathcal{D}
\end{array}%
\right.
\end{eqnarray}%

The Holevo function (\ref{lower2}) assumes different forms, depending on the
equivalence class which the Gaussian channel belongs to.
However, interestingly enough, the value of its maximum is a function of the
channel itself and is not uniquely determined by the property of belonging 
to a given class.

This is reminiscent of other problems in quantum information science 
for which the orbits of the action of the relevant unitary group do not 
provide a complete characterization of the theoretical framework \cite{Schmidt}.


\ack
The authors acknowledge fruitful scientific discussions with 
Ra\'{u}l Garc\'{\i}a-Patr\'{o}n and Oleg V. Pilyavets, and
valuable comments from Lorenzo Maccone and Vittorio Giovannetti.
The research leading to these results has received
funding from the European Commission's seventh Framework Programme
(FP7/2007-2013) under grant agreement HIP, number FP7-ICT-221889.
S.P.\ was supported by a Marie Curie
Action within the 6th European Community Framework
Programme.

\end{document}